\documentclass[%
      aps,
      prd,
      floatfix,
      preprintnumbers,
      preprint,
      showpacs,
      nofootinbib,
      tightenlines,
      amssymb,
      amsmath
]{revtex4-1}

\usepackage{url}
\usepackage{ifpdf}
\ifpdf
	\usepackage{cmap} 
	\usepackage[pdftex]{color,graphicx}
	\usepackage{epstopdf} 
\else
	\usepackage{color,graphicx}
\fi
\usepackage{bm}
\usepackage[dvipsnames]{xcolor}
\usepackage{hyperref}
\hypersetup{%
colorlinks=true,%
citecolor=Blue,%
filecolor=Black,%
linkcolor=Blue,%
urlcolor=Violet
}
\ifpdf
\usepackage[activate={true,nocompatibility},final,tracking=true,kerning=true,spacing=true,factor=1100,stretch=10,shrink=10]{microtype}
\fi

\newcommand{\eq}[1]{Eq.(\ref{#1})}

\begin{document}
\title{
Probing Free Nucleons with (Anti)neutrinos
}

\author{R.~Petti}
\email[]{Roberto.Petti@cern.ch}
\affiliation{Department of Physics and Astronomy, University of South Carolina, Columbia, South Carolina 29208, USA}


\begin{abstract}

We discuss a method to study free protons and neutrons using $\nu(\bar \nu)$-hydrogen (H) Charged Current (CC) inelastic interactions, together with 
various precision tests of the isospin (charge) symmetry using $\nu$ and $\bar \nu$ CC interactions on both H and nuclear targets.  
Probing free nucleons with (anti)neutrinos provides information about their partonic structure, as well as a crucial input for the modeling of $\nu(\bar \nu)$-nucleus (A) interactions. 
Such measurements can also represent a tool to address some of the limitations of accelerator-based neutrino scattering experiments on nuclear targets, 
originating from the combined effect of the unknown (anti)neutrino energy and of the nuclear smearing. 
We also discuss a method to impose constraints on nuclear effects and calibrate the (anti)neutrino energy scale in 
$\nu(\bar \nu)$-A interactions, which are two outstanding systematic uncertainties affecting present and future long-baseline neutrino experiments.

\end{abstract}


\maketitle

\section{Introduction}
\label{sec:intro}

Elucidating the internal structure of protons and neutrons and how quarks and gluons contribute to their momentum, spin, and other intrinsic 
properties like mass and magnetic moment is a crucial step to understand the nature of visible matter. 
Free protons have been extensively studied using a number of processes at large momentum transfer, including electron and muon deep inelastic 
scattering (DIS), lepton-pair production (Drell-Yan process), jet production, and W and Z boson production in $pp$ collisions. 
New dedicated fixed target~\cite{Dudek:2012vr} and collider~\cite{AbdulKhalek:2021gbh} experimental programs are expected to further 
improve our understanding of the proton structure using the electron probe.  
The neutrino and antineutrino probe can potentially add complementary information about free protons (hydrogen) thanks to their unique properties  
including being sensitive only to weak interactions, natural polarization, and a complete flavor separation 
through the CC process. 

Our understanding of the structure of free neutrons is still relatively limited compared to protons, as no direct probe is experimentally feasible. 
Most of our knowledge is obtained by comparing data from the processes described above in proton and deuterium targets, the latter being considered 
as an effective neutron target. The underlying assumption is that the deuteron can be approximated by the sum of a quasi-free proton and a 
quasi-free neutron since it is a weakly bound system. However, both experimental measurements and nuclear models 
indicate that nuclear effects in the deuteron are non-negligible~\cite{Griffioen:2015hxa,Alekhin:2022tip,Alekhin:2017fpf,Kulagin:2004ie,Accardi:2011fa,Accardi:2016qay} 
and strongly depend on the Bjorken $x$ and the momentum transfer $Q^2$, 
adding substantial uncertainties in the study of the neutron structure. Different approaches have been pursued using the electron probe to minimize the impact of 
nuclear effects~\cite{CLAS:2014jvt,MARATHON:2021vqu}. To this end, the combined use of neutrinos and antineutrinos off a 
hydrogen target can potentially offer a more direct access to the free neutron structure by exploiting the flavor selection of the CC process and the isospin symmetry. 

The availability of new precision measurements from (anti)neutrino-hydrogen interactions would concurrently provide  
a valuable tool to address some of the limitations of neutrino scattering experiments using nuclear targets~\cite{Petti:2019asx}. 
The energy of the projectile (anti)neutrino is unknown on an event-by-event basis and can vary over a broad range in conventional wide band neutrino beams. 
For this reason neutrino experiments have been affected by relatively large systematic uncertainties on the knowledge of the (anti)neutrino flux. 
The initial momentum of the target nucleon within the nucleus is also unknown and hadrons produced in the primary interactions 
can undergo an additional unknown modification as they can be absorbed or re-interact within the nucleus (final state interactions). 
Neutrino scattering experiments have to infer the (anti)neutrino energy from the detected final state particles emerging from the 
nucleus, which are affected by a substantial nuclear smearing and related systematic uncertainties. 
This procedure typically implies model corrections depending upon a number of parameters, often empirically tuned with the observed kinematic distributions. 
In order to make the problem tractable a target of well known energy is required in the absence of monochromatic (anti)neutrino beams. 
We could therefore argue that the availability of a hydrogen target -- the only hadron target of known energy -- 
is necessary to go beyond the precision level of existing neutrino scattering experiments~\cite{Petti:2019asx}. Using exclusive $\nu_\mu$ single pion and 
$\bar \nu_\mu$ quasi-elastic processes on hydrogen at small energy transfer allows the determination of the shape of the $\nu_\mu$ and $\bar \nu_\mu$ fluxes 
as a function of energy with an accuracy better than 1\% in conventional wide-band neutrino beams~\cite{Duyang:2019prb}.
Furthermore, a direct comparison of CC interactions on H and on nuclear targets within the same detector can provide a calibration 
of the reconstructed neutrino energy scale and a reduction of the corresponding systematic uncertainties in data collected from nuclear targets.

In spite of their extreme relevance, the available data from $\nu(\bar \nu)$-H interactions is rather sparse and limited to 
the early bubble chamber experiments ANL 12-foot~\cite{Barish:1977qk}, BNL 7-foot~\cite{Fanourakis:1980si}, FNAL E31~\cite{Derrick:1981zb} 
and E45~\cite{Bell:1978ta}, CERN WA21~\cite{WA21:1990dkk} and WA24~\cite{BEBCTSTNeutrino:1983vyc}. The largest samples correspond 
to a total of about 16k $\nu$-H and 9k $\bar \nu$-H CC interactions -- collected some 40 years ago -- with the bulk of bubble chamber data 
being actually taken on deuterium or heavier targets. Since then safety requirements related to the underground operation and 
practical considerations favoring electronic detectors have prevented new measurements. 

From the discussion above it is clear that the existing $\nu(\bar \nu)$-H data are inadequate for the needs of future neutrino 
scattering experiments. These latter would require new high resolution samples with statistics commensurate to the one 
expected to be collected from nuclear targets, corresponding roughly to an increase of at least two orders of magnitude 
with respect to existing $\nu(\bar \nu)$-H data. Considering the high intensity of modern (anti)neutrino beams, 
such samples can be obtained with a fiducial mass of H close to 1 ton~\footnote{For comparison, the WA21 experiment 
used a bubble chamber with a fiducial volume of 19 m$^3$ corresponding to a liquid hydrogen mass of about 1.3 tons~\cite{WA21:1990dkk}.}.  
This value implies the use of H in liquid form or within solid compounds for a realistic detector size. 

In this paper we discuss the use of $\nu(\bar \nu)$-H interactions to study free protons and neutrons, 
as well as to calibrate the neutrino energy scale in CC interactions with nuclei, 
focusing on the ``solid" hydrogen concept we recently proposed~\cite{Petti:2019asx}. 
Sec.~\ref{sec:proton} briefly summarizes the method to obtain $\nu(\bar \nu)$-H interactions, 
while Sec.~\ref{sec:neutron} describes the use of the isospin symmetry to determine the partonic structure of free neutrons. 
Section~\ref{sec:nbound} focuses on the case of bound nucleons in deuterium and other nuclei. 
In Sec.~\ref{sec:isospin}  we discuss various tests of the isospin (charge) symmetry exploiting the same detector concept. 
In Sec.~\ref{sec:xsec} we describe the application to free nucleon cross-sections. In Sec.~\ref{sec:nuclear} we discuss how 
$\nu(\bar \nu)$-H interactions can be used to constrain the nuclear smearing in nuclear targets and in 
Sec.~\ref{sec:calibration} we present a method to calibrate the neutrino energy scale for interactions on nuclei.

\section{Free Proton Target} 
\label{sec:proton} 

With the ``solid" hydrogen concept $\nu(\bar \nu)$ interactions on free protons are obtained by subtracting measurements on dedicated 
graphite (C) targets from those on polypropylene (CH$_2$) targets. A large number of thin planes -- each typically 1-2\% of radiation length -- of both materials 
with comparable thickness are alternated and dispersed throughout a Straw Tube Tracker (STT) of negligible mass in order to guarantee the same acceptance 
to final state particles produced in (anti)neutrino interactions. The STT allows to minimize the thickness of individual active layers 
-- made of four straw planes -- and to approximate the ideal case of a pure target detector -- the CH$_2$ and C targets corresponding to about 97\% of the total mass -- 
while keeping the total thickness of the stack comparable to one radiation length. 
Each target plane can be removed or replaced with different materials during data taking, providing a flexible target configuration.  
We emphasize that the STT is an essential element of the ``solid" hydrogen concept since it is designed to provide a control of the configuration, 
chemical composition, and mass of the neutrino targets similar to electron scattering experiments. 
The technique is conceived for a model-independent subtraction of the C background by using the data from the measurements on the graphite targets, 
automatically including all types of interactions, as well as detector effects, relevant for the H selection~\footnote{The approach is conceptually similar 
to what is done in electron scattering experiments, in which a cryogenic tank is filled (the CH$_2$ targets) with 
a liquid H$_2$ target and dedicated runs with the empty tank (the graphite targets) are taken for background subtraction.}.  

The low average density of the detector -- similar to that of liquid deuterium $\rho \sim 0.17$ g/cm$^3$ -- and the overall dimensions comparable 
to one radiation length allow an accurate reconstruction of the four-momenta of the visible final state particles, as well as of the event kinematics in a plane transverse 
to the beam direction~\footnote{The detector is based on a concept similar to the NOMAD experiment, which was explicitly designed to exploit the 
transverse plane kinematics for event selection~\cite{Astier:2001yj}.}. The momentum scale can be calibrated to about 0.2\% using reconstructed 
$K_0 \to \pi^+\pi^-$ and $\Lambda \to p \pi^-$ decays~\cite{Wu:2007ab,Duyang:2019prb}.
For events with at least two reconstructed charged tracks the primary interaction 
vertex can be associated to the correct target layer with an uncertainty less than 0.1\% 
thanks to a vertex resolution ($\ll 1$ mm~\cite{Anfreville:2001zi}) much smaller than the target thickness, 
together with the lightness of the tracking straws and the chemical purity of the targets. These latter factors are critical for events with a single reconstructed 
charged track, for which the uncertainty in associating the event to the correct material is given by the 
ratio between the thickness of the straw walls and the one of a single target layer, typically below 0.5\%. 

\begin{figure}[tb]
\begin{center}
\includegraphics[width=0.65\textwidth]{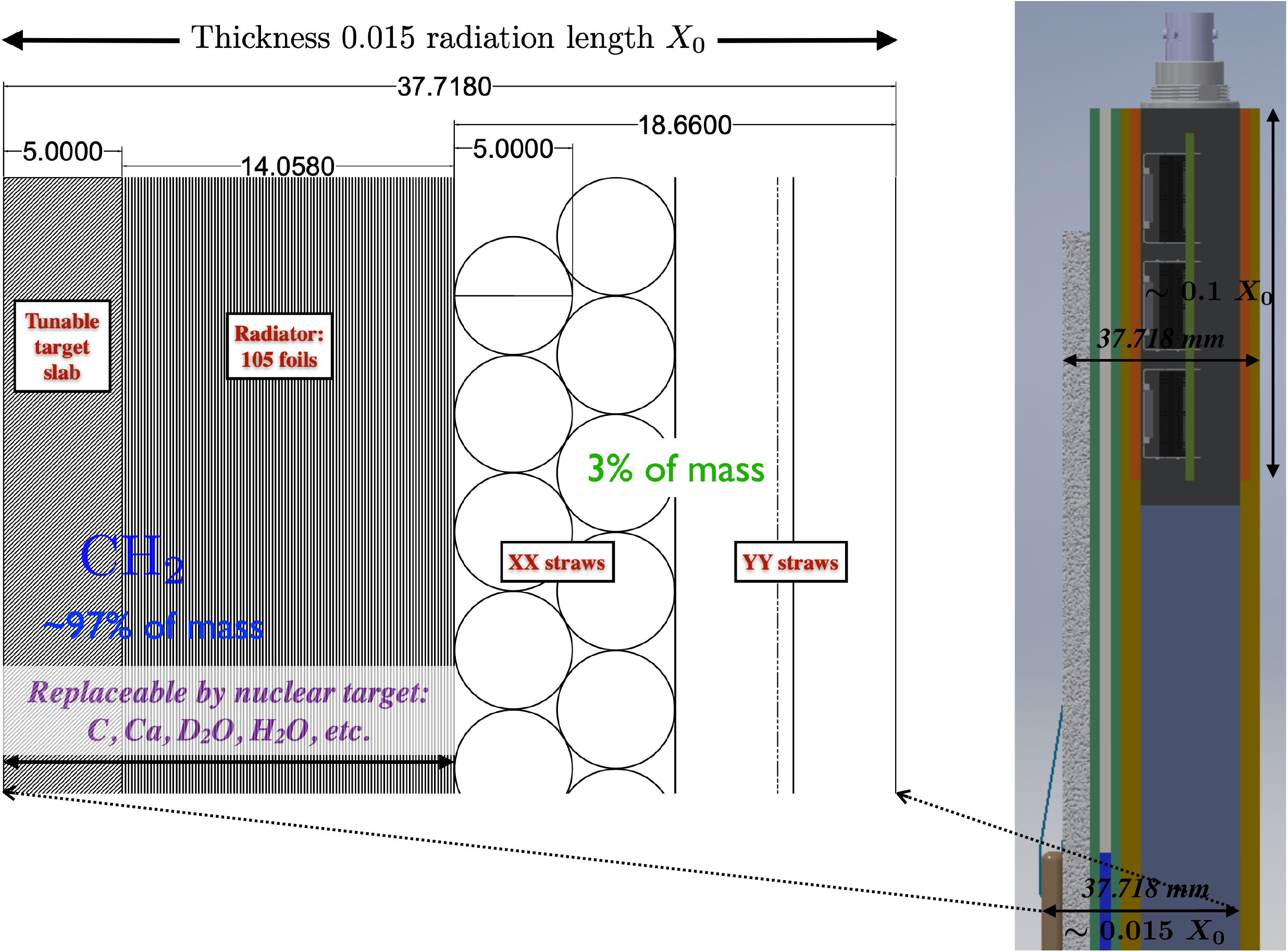}
\end{center}
\caption{Schematic drawing of a STT module allowing a control of the configuration, chemical composition, and mass of the $\nu(\bar \nu)$ target(s) comparable to electron scattering experiments.
}
\label{fig:STT_DesignConcept}
\end{figure}

The subtraction technique described above can be used to select any inclusive and exclusive process in both CC and Neutral Current (NC) $\nu(\bar \nu)$ 
interactions on free protons at the price of an increased statistical uncertainty from the subtraction procedure. 
For CC interactions the dilution factor with respect to a pure H$_2$ target can be drastically reduced with a kinematic analysis based on 
energy-momentum conservation exploiting the excellent resolution of STT~\cite{Duyang:2018lpe}. 
Since the H target is at rest, the CC events are expected to be perfectly balanced in a plane transverse to the beam direction. 
Instead, events from nuclear targets are affected by nuclear effects, resulting in a significant missing transverse momentum and a smearing of the transverse plane kinematics. 
These differences can be exploited for the selection of all available inclusive and exclusive topologies in both $\nu$ and $\bar \nu$ CC interactions on H~\cite{Duyang:2018lpe}, 
increasing the purity of the corresponding H samples in the range 80-95\% before subtraction, depending upon the specific process considered.
While the kinematic analysis can significantly enhance the sensitivity of the measurements, 
we note that the ``solid" hydrogen concept can still be used with no or limited kinematic selection. 

In principle the use of a pure liquid H$_2$ target is still preferable over the subtraction technique if the safety and technical aspects can be 
addressed~\cite{Alvarez-Ruso:2022exy}. 
However, the ``solid" hydrogen concept can be a viable option offering an acceptable approximation which is both safe and relatively inexpensive to implement. 
Typical fiducial masses equivalent to about 10 m$^3$ of liquid H$_2$ can be easily achieved in this way. 
One specific advantage of this technique is the availability of an integrated pure carbon target, as well as the possibility to install a variety of nuclear targets 
within the same detector with comparable acceptance. Some of the implications of such targets are discussed in the following.

\section{Free Neutron Target} 
\label{sec:neutron} 

The absence of a physical neutron target poses the challenge of extracting information about free neutrons without 
relying on model corrections for the sizable nuclear effects on bound nucleons in nuclei. The availability of both $\nu$-H and $\bar \nu$-H CC interactions can perhaps 
offer the most direct solution by exploiting the isospin symmetry in nucleon structure functions $F^{\bar \nu p} = F^{\nu n}$. 
This relationship is expected to be valid only in the limit of neglecting quark mixing and heavy flavor production 
since the isospin is an approximate symmetry of the strong interactions. In order to quantify the deviations introduced by such effects 
in CC weak interactions we can consider the following quantities: 
\begin{eqnarray} 
{\cal R}_2^{\rm p/n} (x,Q^2) & = & \frac{F_2^{\bar \nu p} (x,Q^2) }{F_2^{\nu n} (x,Q^2)} - 1 =  \frac{F_2^{\bar \nu p} (x,Q^2) -F_2^{\nu n} (x,Q^2)}{F_2^{\nu n} (x,Q^2)} \label{eq:Hevt2} \\  
{\cal R}_3^{\rm p/n} (x,Q^2) & = & \frac{xF_3^{\bar \nu p} (x,Q^2)}{xF_3^{\nu n} (x,Q^2)} - 1 =  \frac{xF_3^{\bar \nu p} (x,Q^2) -xF_3^{\nu n} (x,Q^2)}{xF_3^{\nu n} (x,Q^2)} 
\label{eq:Hevt3} 
\end{eqnarray} 
where $F_2^{\bar \nu p} (xF_3^{\bar \nu p})$ and $F_2^{\nu n} (xF_3^{\nu n})$ refer to the $F_2 (xF_3)$ structure functions for the CC processes  
$\bar \nu_\mu p \to \mu^+ X$ and $\nu_\mu n \to \mu^- X$ on a free proton and a free neutron, respectively. 
The quantities ${\cal R}_2^{\rm p/n}$ and ${\cal R}_3^{\rm p/n}$ are not directly measurable experimentally but represent the size of the corrections 
to be applied on the free neutron structure functions $F_2^{\nu n}$ and $xF_3^{\nu n}$ extracted from $\bar \nu$-H interactions on the basis of isospin symmetry. 

\begin{figure}[p]
\begin{center}
\includegraphics[width=0.50\textwidth]{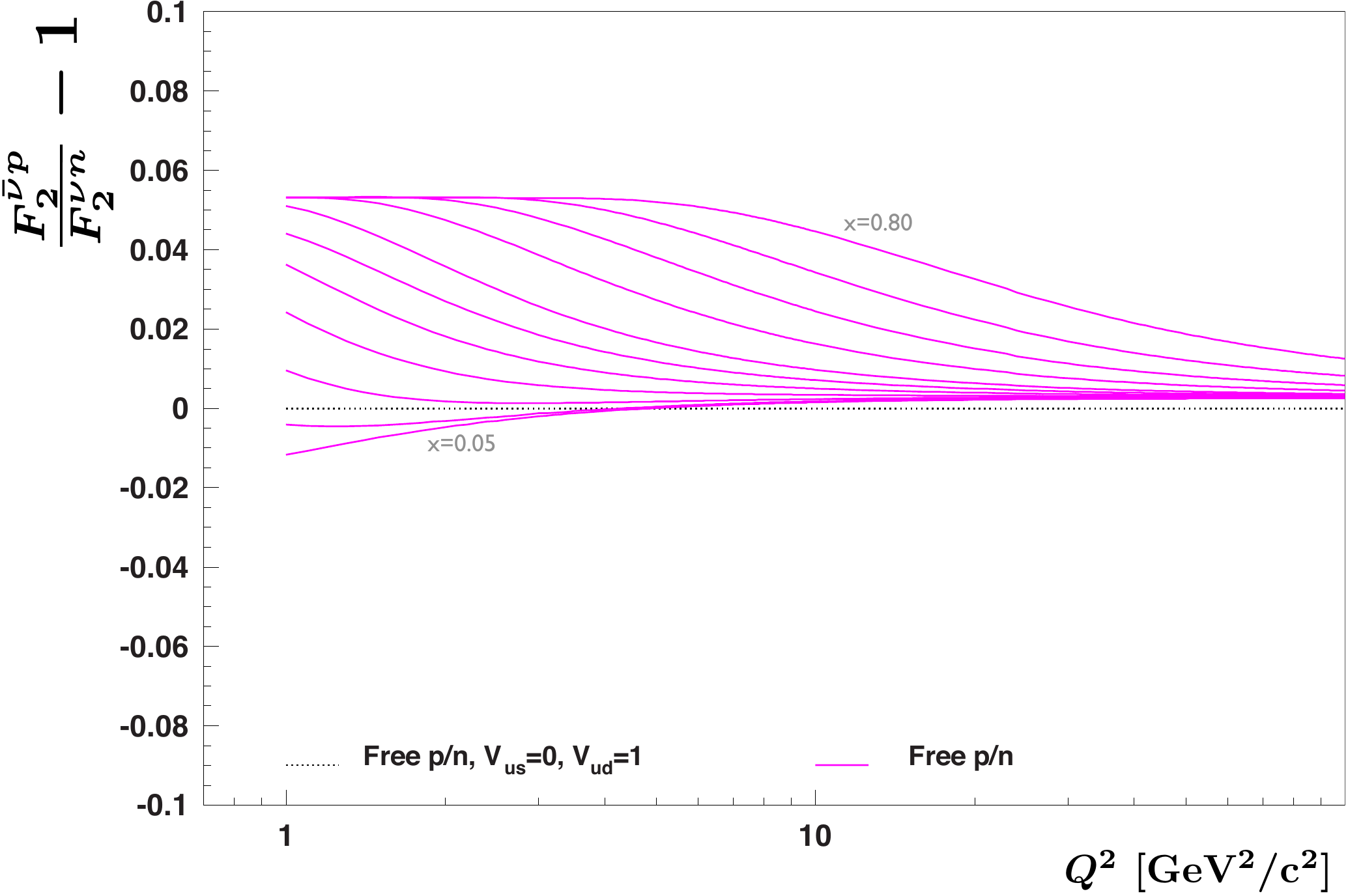}\hspace*{0.10cm}\includegraphics[width=0.50\textwidth]{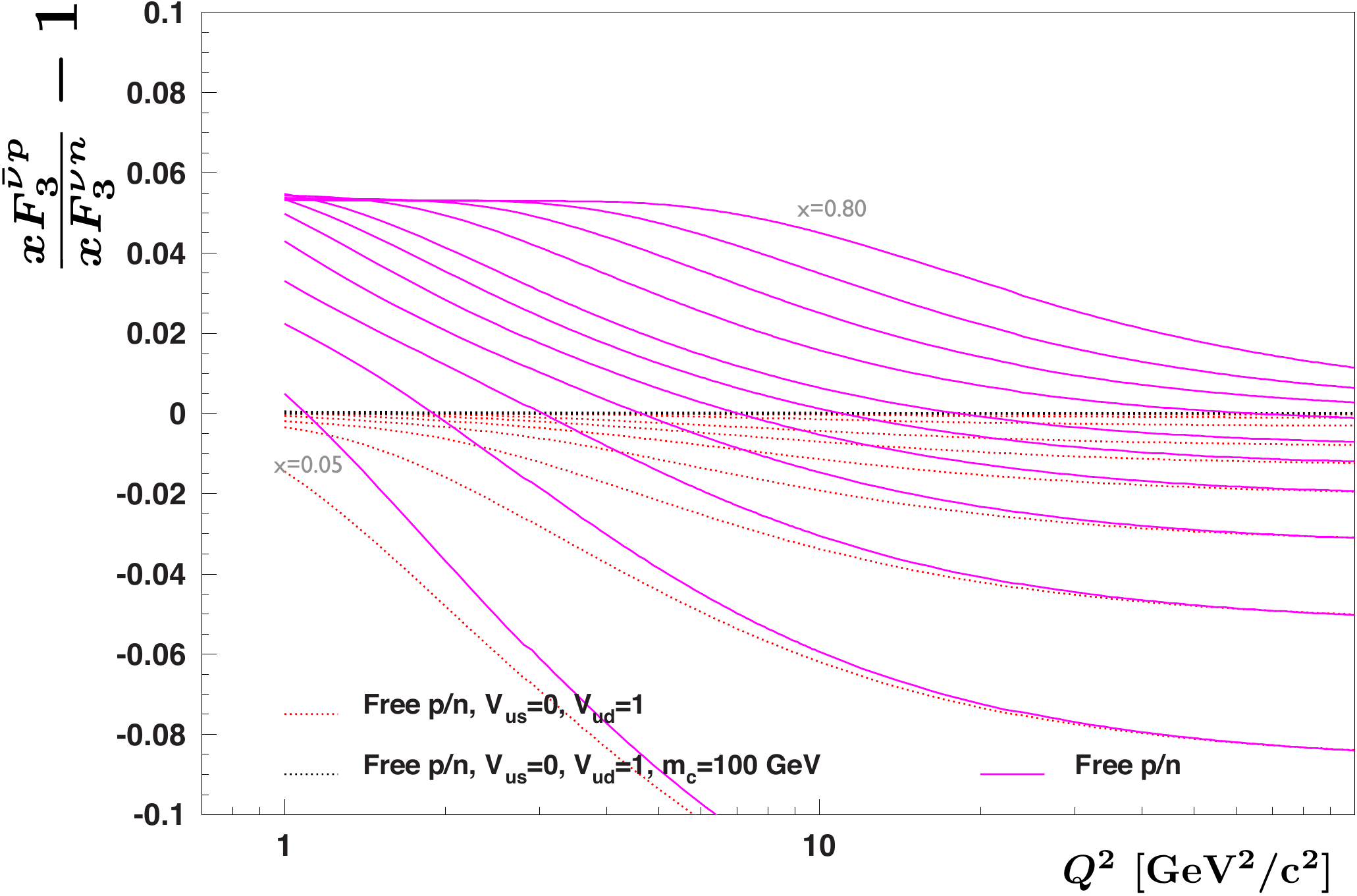}
\end{center}
\caption{Left plot: ratio ${\cal R}_2^{\rm p/n}$ for the $F_2$ structure functions of free nucleons as a function of the momentum transfer $Q^2$ (solid lines). 
The effect of the quark mixing is also shown for comparison (dotted lines). 
The various curves correspond (from bottom to top) to values of $x=0.05,0.1,0.15,0.2,0.25,0.3,0.35,0.4,0.5,0.6,0.7,0.8$. 
Right plot:  same notations as the previous plot for the ratio ${\cal R}_3^{\rm p/n}$ for the $xF_3$ structure functions of free nucleons as a function of the momentum transfer $Q^2$. The effect of both the quark mixing and of the mass of the charm quark are also shown for comparisons. 
See text for details.
}
\label{fig:isospin-H}
\end{figure}

We calculated ${\cal R}_2^{\rm p/n}$ and ${\cal R}_3^{\rm p/n}$ in the QCD factorization scheme with three fixed flavours
at the NNLO approximation in the strong coupling constant, using the results of the global QCD analysis of Ref.~\cite{Alekhin:2007fh} for the 
structure functions of free nucleons and including target mass corrections~\cite{Georgi:1976ve}. 
Since most determinations of the strange sea charge asymmetry of the nucleon are consistent with 
zero~\cite{Alekhin:2008mb,Martin:2009iq,Lai:2007dq,Ball:2009mk,NuTeV:2007uwm}, we assumed $s(x)=\bar s(x)$. 
The results are shown in Fig.~\ref{fig:isospin-H} as a function of the momentum transfer $Q^2$ for different values of $x$. 
The deviation from zero introduced by the quark mixing on ${\cal R}_2^{\rm p/n}$ is negligible at small $x$ and progressively grows up to 
about 5\% at larger $x$ values for small values of $Q^2$. At large $Q^2$ the corrections are smaller than 1\% in the entire $x$ range available. 
Since the isospin symmetry $u_{p(n)} = d_{n(p)}$ was assumed in the underlying QCD analysis~\cite{Alekhin:2007fh} all deviations vanish by setting $V_{us}=0$ and 
$V_{ud}=1$. 
We note that the sign of the expected corrections on ${\cal R}_2^{\rm p/n}$ from the quark mixing is essentially positive. 
The same applies to ${\cal R}_3^{\rm p/n}$ as well, as can be seen from the right plot in Fig.~\ref{fig:isospin-H}. 
However, in the latter case an additional negative contribution from charm quark production is present, as the difference 
$xF_3^{\bar \nu p} - xF_3^{\nu n}$ in \eq{eq:Hevt3} is directly sensitive to the strangeness content of the nucleon in addition to 
isospin effects. For this reason setting $V_{us}=0$ and $V_{ud}=1$ results in negative values of ${\cal R}_3^{\rm p/n}$, which becomes zero only 
below threshold for charm quark production at small $Q^2$. This result can be verified with an exceedingly large value for the mass 
of the charm quark $m_c=100$ GeV/c$^2$ (Fig.~\ref{fig:isospin-H}) and  explains why the corresponding correction 
increases with $Q^2$, contrary to the one associated to the quark mixing. 

We note that the impact of the corrections introduced by the quark mixing is reduced by the fact that $V_{ud}$ is currently known 
with an accuracy of about $2\times 10^{-4}$ and $V_{us}$ with an accuracy of about $3 \times 10^{-3}$~\cite{ParticleDataGroup:2020ssz}. 
The existing knowledge of the charm quark mass $m_c$  (about 2\%)~\cite{ParticleDataGroup:2020ssz} and of the 
strange sea content of the nucleon~\cite{Alekhin:2014sya,Alekhin:2018dbs} 
are also adequate to the accuracy required for the calculation of the corresponding corrections to ${\cal R}_3^{\rm p/n}$. 
This precision can be further improved with a dedicated analysis of exclusive charm production in the large samples of 
$\nu$ and $\bar \nu$ CC interactions collected by the detector being considered. In general, 
all required corrections to ${\cal R}_2^{\rm p/n}$ and ${\cal R}_3^{\rm p/n}$ are relatively small and, most importantly, 
entirely related to the partonic structure of free nucleons. 

Since the hadronization process is controlled by the strong interaction the isospin symmetry can also provide valuable information about exclusive final states 
produced in CC interactions with free neutrons. To this end, we can use exclusive processes in $\bar \nu$-H CC interactions and 
replace the relevant hadrons by the corresponding isospin-rotated states: 
\begin{eqnarray} 
p & \longleftrightarrow & n  \nonumber  \\ 
\pi^+ & \longleftrightarrow &  \pi^-  \nonumber 
\end{eqnarray} 
and similar relations for other detected particles. 
While $\pi^+$ and $\pi^-$ have a similar experimental signature, the replacement $p \leftrightarrow n$ requires to apply an acceptance correction 
taking into account the different detection efficiency of protons and neutrons. The proton reconstruction can be accurately calibrated with the 
large sample of $\Lambda \to p \pi^-$ decays available and the absolute neutron detection efficiency can be calibrated with dedicated 
testbeam exposures of the relevant detector components~\cite{Duyang:2019prb}. 

\begin{figure}[p]
\begin{center}
\includegraphics[width=0.50\textwidth]{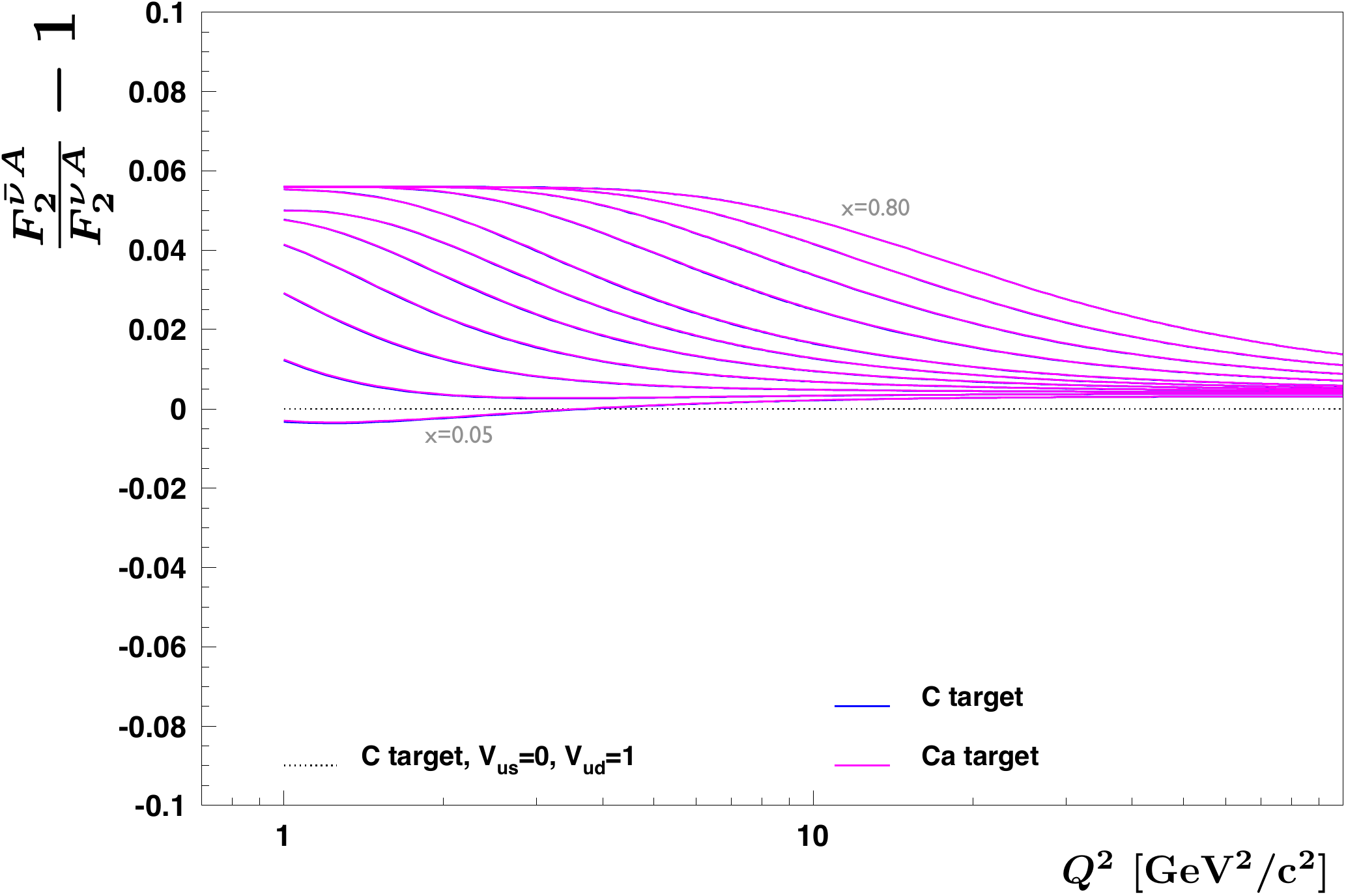}\hspace*{0.10cm}\includegraphics[width=0.50\textwidth]{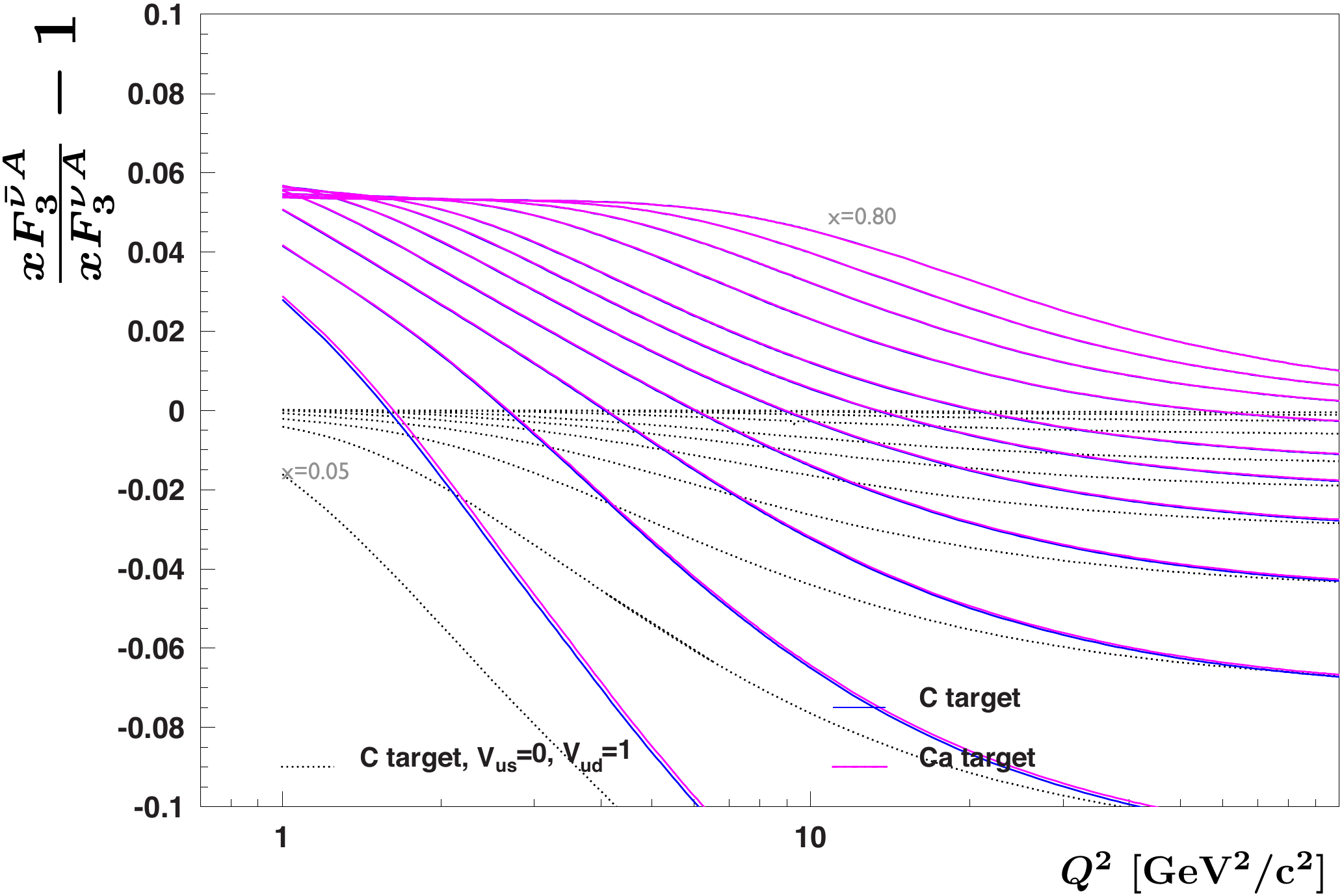}
\end{center}
\caption{Same notations as Fig.~\ref{fig:isospin-H} for the ratios ${\cal R}_2^{\rm A}$ (left plot) and ${\cal R}_3^{\rm A}$ (right plot) in the C and Ca isoscalar nuclei. 
See text for details. 
}
\label{fig:isospin-CaC}
\end{figure}

\section{Bound Nucleon Targets} 
\label{sec:nbound} 

Although nuclear effects in deuterium (D) are not negligible and can introduce significant uncertainties in the extraction of the 
free neutron structure functions~\cite{Griffioen:2015hxa,Alekhin:2022tip,Alekhin:2017fpf,Kulagin:2004ie,Accardi:2011fa,Accardi:2016qay}, the availability of 
such a light nuclear target can still provide relevant information on the bound n-p system. 
The same considerations about detectors based on liquid H$_2$ apply to the case of liquid D$_2$. 
We can exploit the precise control of the targets offered by STT (Sec.~\ref{sec:proton}) to obtain $\nu$ and $\bar \nu$ interactions off the bound 
neutron in the deuteron from a subtraction between measurements on heavy water (D$_2$O) and ordinary water (H$_2$O). 
Both water targets must have identical thickness -- roughly $\leq 5\%$ of radiation length -- and be enclosed into identical plastic shells hermetically sealed. 
Several planes can be integrated into the detector by replacing some of the default CH$_2$ targets. 
The distribution of the generic kinematic variables $\vec x \equiv (x_1, x_2, ....., x_n)$ in $\nu(\bar \nu)$-D CC interactions can be obtained as: 
\begin{equation} 
N_{\rm D}(\vec x) \equiv \frac{N_{\rm D_2 O} (\vec x) - N_{\rm H_2 O} (\vec x)}{2} + \frac{M_{\rm H_2/H_2O}}{2 M_{\rm H}} N_{\rm H} (\vec x)
\label{eq:Devt} 
\end{equation} 
where $N_{\rm D_2 O}$ and $N_{\rm H_2 O}$ are the data from the D$_2$O and H$_2$O targets, which are designed to have the same total 
mass of oxygen, and $N_{\rm H}$ is the number of events from the ``solid" hydrogen target. The interactions from this latter are normalized 
by the ratio between the total fiducial masses of H within the ordinary water and the H targets, $M_{\rm H_2/H_2O}/M_{\rm H}$. 
The simple subtraction between the two water targets provides instead interactions on free neutrons supplemented by the total nuclear 
effects in the deuteron (${\rm D}-n-p$). 
Comparing the measurements of the bound nucleon structure functions $F_{2,3}^{\rm \nu D}$ with the ones of the 
free proton $F_{2,3}^{\nu p}$ and the free neutron $F_{2,3}^{\nu n}$ based on the isospin symmetry $\bar \nu p \leftrightarrow \nu n$ 
(Sec.~\ref{sec:neutron}) can potentially provide the first direct measurement of nuclear effects in the deuteron with the ratio 
$R_{2,3}^{\rm \nu D} = F_{2,3}^{\rm \nu D} / (F_{2,3}^{\nu p} + F_{2,3}^{\nu n})$. 
Similar measurements can be performed with the various nuclear targets which can be integrated within the STT (Sec.~\ref{sec:proton}) 
by considering the ratio $R_{2,3}^{\rm \nu A} = F_{2,3}^{\rm \nu A} / \left[ Z F_{2,3}^{\nu p} + (A-Z) F_{2,3}^{\nu n} \right]$ 
for a a generic target nucleus $A$ with $Z$ protons and $(N-Z)$ neutrons. Using a combination of both isoscalar (e.g. the integrated C target)
and non-isoscalar nuclear targets can provide valuable insights on the physics mechanisms responsible of the nuclear modifications 
of the nucleon properties.

\section{Tests of Isospin Symmetry} 
\label{sec:isospin} 

We have discussed how the availability of both $\nu$-H and $\bar \nu$-H CC interactions can provide direct information on the partonic structure of the free neutron.
Since the method relies on the isospin symmetry we need to include the possibility to verify the limit of validity of such an assumption into the 
corresponding experimental program. Precision tests of the isospin (charge) symmetry also represent valuable physics measurements per se as they 
can shed light on how QCD works in its non-perturbative regime.

The Adler sum rule~\cite{Adler:1964yx,Adler:2009dw} relates the integral of the isovector combination $F_2^{\bar \nu} - F_2^{\nu}$ over $x$  
to the isospin of the target: 
\begin{equation} 
S_{\rm A} (Q^2) = \int_0^1 \frac{dx}{2x} \left[ F_2^{\bar \nu } (x,Q^2) - F_2^{\nu } (x,Q^2) \right] = 2 I_z 
\label{eq:Adler} 
\end{equation} 
where $I_z$ is the projection of the target isospin vector on the quantization axis (z axis). The Adler integral represents an exact sum rule derived from current algebra. 
For a H target (free proton) $S_{\rm A}^p = 1$, while for a generic nucleus with Z protons and N neutrons $S_{\rm A}^A = (Z-N)/A = \beta$~\cite{Kulagin:2007ju}. 
The \eq{eq:Adler} survives the strong interaction because of the conservation of the vector current, but it neglects the effects of the non-conservation of the 
axial current, as well as quark mixing and heavy flavor production. A precision measurement of $S_{\rm A}^p$ as a function of $Q^2$ using 
$\nu$-H and $\bar \nu$-H CC interactions is directly sensitive to violations of the isospin (charge) symmetry in free nucleons. 
The only existing measurement was performed by BEBC with less than 10k events~\cite{Allasia:1985hw}. 

\newpage
For isoscalar nuclei the isospin symmetry implies $F^{\bar \nu A} = F^{\nu A}$ and deviations from this behavior can be studied with the following 
quantities: 
\begin{eqnarray} 
{\cal R}_2^{\rm A} (x,Q^2) & = & \frac{F_2^{\bar \nu A} (x,Q^2)}{F_2^{\nu A} (x,Q^2)} - 1 =  \frac{F_2^{\bar \nu A} (x,Q^2) -F_2^{\nu A} (x,Q^2)}{F_2^{\nu A} (x,Q^2)} \\  
{\cal R}_3^{\rm A} (x,Q^2) & = & \frac{xF_3^{\bar \nu A} (x,Q^2)}{xF_3^{\nu A} (x,Q^2)} - 1 =  \frac{xF_3^{\bar \nu A} (x,Q^2) -xF_3^{\nu A} (x,Q^2)}{xF_3^{\nu A} (x,Q^2)} 
\label{eq:Hevt} 
\end{eqnarray} 
which represent the analogy of \eq{eq:Hevt2} and \eq{eq:Hevt3} for nuclear targets.  Quark mixing and heavy flavor production are expected to 
introduce small non vanishing contributions to ${\cal R}_2^{\rm A}$ and ${\cal R}_3^{\rm A}$, similarly to the case of free nucleons discussed in Sec.~\ref{sec:neutron}.
Isoscalar nuclear targets offer an excellent tool for precision tests of the isospin (charge) symmetry by measuring the deviations from zero~\footnote{We expect small 
non-zero values from the difference between the quark masses and from QED corrections.} 
of ${\cal R}_2^{\rm A}$ and ${\cal R}_3^{\rm A}$ as a function of $x$ and $Q^2$. 
To this end, we can exploit the pure C (graphite) target which is an essential element of the ``solid" hydrogen technique to obtain 
$\nu (\bar \nu)$-H interactions (Sec.~\ref{sec:proton}). The isotopic content expected for a standard C target is about 98.9\% of the isoscalar ${}^{12}$C 
and about 1.1\% of ${}^{13}$C, resulting on average in an isovector component $\beta \sim -9 \times 10^{-4}$

\begin{figure}[tb]
\begin{center}
\includegraphics[width=0.50\textwidth]{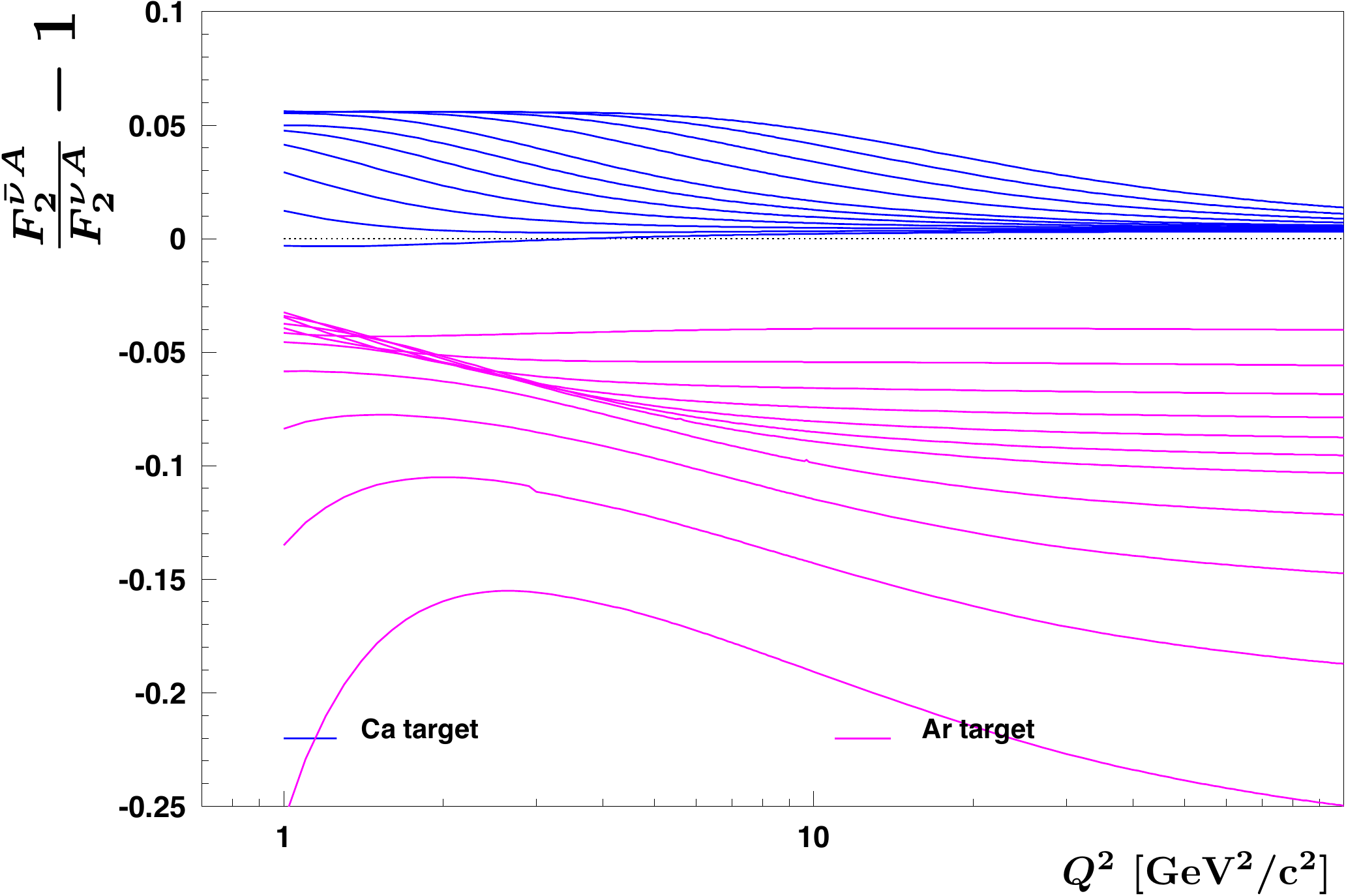}\hspace*{0.10cm}\includegraphics[width=0.50\textwidth]{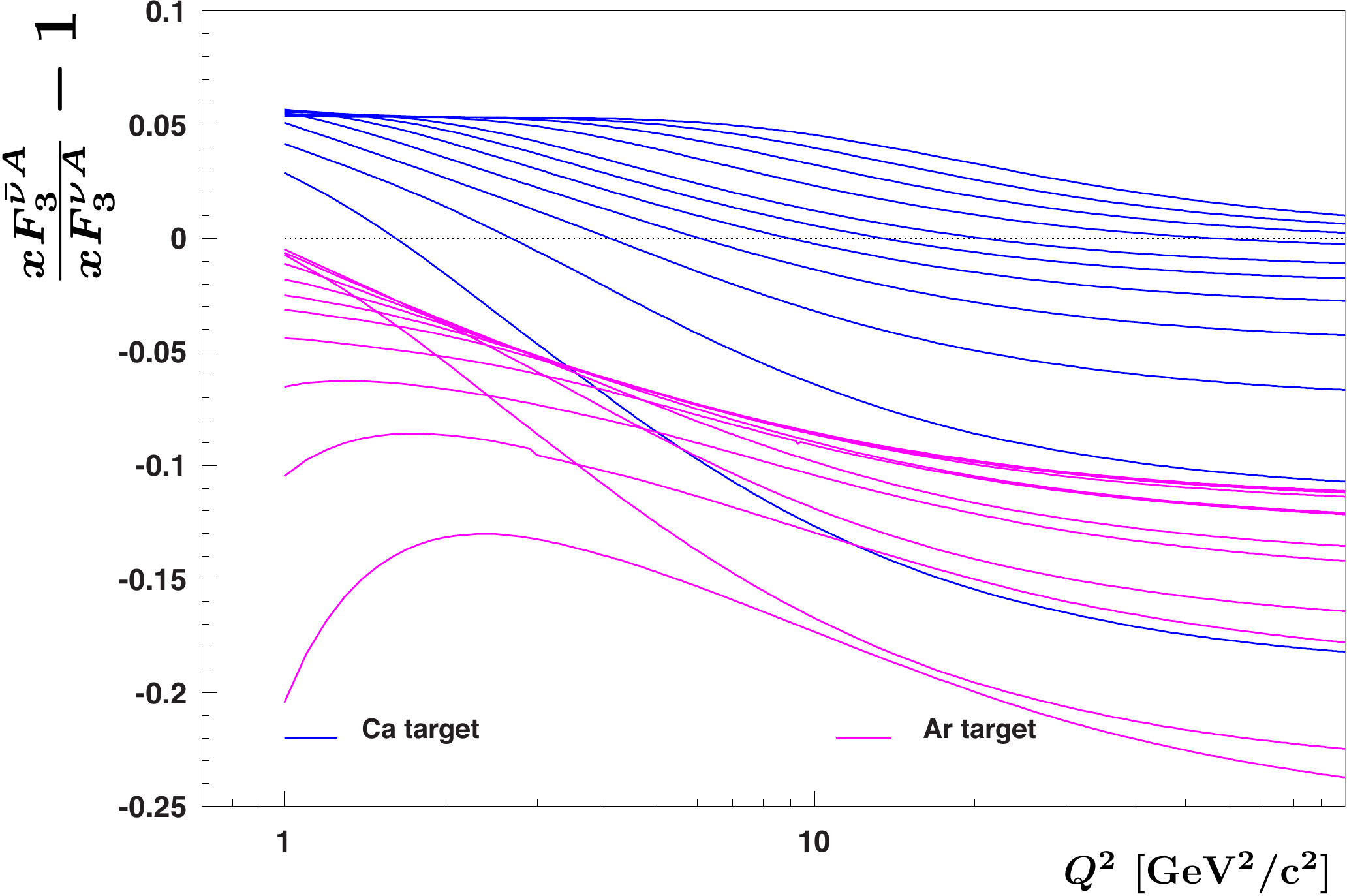}
\end{center}
\caption{Same notations as Fig.~\ref{fig:isospin-CaC} for the ratios ${\cal R}_2^{\rm A}$ (left plot) and ${\cal R}_3^{\rm A}$ (right plot) in the $A=40$ nuclei. 
Results for the isoscalar Ca target are compared with the Ar target characterized by a significant isovector component. 
See text for details. 
}
\label{fig:isospin-ArCa}
\end{figure}

We calculated ${\cal R}_2^{\rm A}$ and ${\cal R}_3^{\rm A}$ for various nuclei using 
the nuclear model of Refs.~\cite{Kulagin:2004ie,Kulagin:2007ju,Kulagin:2014vsa} which includes 
the smearing with the energy-momentum distribution 
of bound nucleons (Fermi motion and binding), the off-shell correction to bound nucleon structure functions, 
the contributions from meson exchange currents and the propagation of the hadronic component of the 
virtual intermediate boson in the nuclear environment.
The underlying nucleon structure functions are the same described in Sec.~\ref{sec:neutron}. 
This model has been successfully used to quantitatively explain the 
observed $x$, $Q^2$ and $A$ dependence of 
the nuclear DIS data in a wide range of targets from the deuteron to ${}^{207}$Pb~\cite{Kulagin:2004ie,Kulagin:2010gd,Alekhin:2017fpf,MARATHON:2021vqu}, 
the magnitude, the $x$ and mass dependence of the nuclear Drell-Yan (DY) data~\cite{Kulagin:2014vsa}, as well as the data on  
the differential cross sections and asymmetries for $W^\pm,Z$ production in $p+{\rm Pb}$ collisions 
at the LHC~\cite{Ru:2016wfx}.  
Results are shown in Fig.~\ref{fig:isospin-CaC} and Fig.~\ref{fig:isospin-ArCa}. 

A comparison between Fig.~\ref{fig:isospin-H} and Fig.~\ref{fig:isospin-CaC} indicates that nuclear effects in isoscalar nuclei do not alter significantly the 
behavior of ${\cal R}_2^{\rm A}$ and ${\cal R}_3^{\rm A}$ and that these latter follow very closely ${\cal R}_2^{\rm p/n}$ and ${\cal R}_3^{\rm p/n}$  
in free nucleons. Measurements from the pure C target can then be directly used to validate the small correction factors applied in the determination of the free neutron 
structure functions from $\nu$-H and $\bar \nu$-H CC interactions. The role of the C target in the ``solid" hydrogen technique thus extends 
beyond the simple background subtraction. 

The values of ${\cal R}_2^{\rm A}$ and ${\cal R}_3^{\rm A}$ measured with the C target can also be used to search for direct violations of 
the isospin (charge) symmetry from the observation of deviations with respect to the behavior shown in Fig.~\ref{fig:isospin-CaC}. 
In the case of ${\cal R}_2^{\rm A}$ such deviations could also be translated into anomalous values of $V_{ud}$ and $V_{us}$ controlling 
the corresponding quark mixing. As illustrated in Fig.~\ref{fig:isospin-CaC} the sensitivity to such effects is maximal for negative 
values of ${\cal R}_2^{\rm A}$ and, in general, increases with $Q^2$. For values of the momentum transfer above the charm production 
threshold ${\cal R}_3^{\rm A}$ is also sensitive to the value of the charm quark mass $m_c$ and to the strange quark content of the nucleons. 
For this reason a combined analysis of charm production in CC interactions is required to achieve the ultimate 
sensitivity on potential violations of the isospin (charge) symmetry. 

In case anomalous deviations from the expected values of ${\cal R}_2^{\rm A}$ and ${\cal R}_3^{\rm A}$ are observed from the C target, an independent 
measurement using a different isoscalar nucleus would be required to verify that the potential violations of the isospin (charge) symmetry are not 
introduced by nuclear modifications. To this end, thin solid Ca targets~\footnote{The Ca targets will have to be encapsulated and protected from the environment.} 
could be integrated into the detector in place of some of the standard CH$_2$ targets. 
The isotopic content expected for a standard Ca target is about 96.9\% of the isoscalar ${}^{40}$Ca, 2.1\% of  ${}^{44}$Ca, 0.65\% of ${}^{42}$Ca, 
0.2\% of ${}^{48}$Ca, and 0.14\% of ${}^{43}$Ca, resulting on average in an isovector component $\beta \sim -2.6 \times 10^{-3}$.  
As shown in Fig.~\ref{fig:isospin-CaC} both ${\cal R}_2^{\rm A}$ and ${\cal R}_3^{\rm A}$ in Ca follow closely the corresponding values for C. 
Another advantage of a Ca target is that it is characterized by the same $A=40$ as the dominant ($\sim$99.6\%) Ar isotope. 
This latter nucleus has a sizable neutron excess resulting in an average value of $\beta \sim - 0.1$ for a standard Ar target. 
Since ${\cal R}_2^{\rm A}$ and ${\cal R}_3^{\rm A}$ are sensitive to isovector effects, a comparison between such measurements in 
Ca and Ar can explicitly probe the isospin dependence of nuclear effects and help to better understand the structure of the $A=40$ nuclei. 
Results are illustrated in Fig.~\ref{fig:isospin-ArCa}.  
The neutron excess in Ar is responsible of the relatively large negative values for both quantities. 
We note that nuclear modifications to ${\cal R}_2^{\rm A}$ and ${\cal R}_3^{\rm A}$ are significantly larger for non-isoscalar nuclei. 
In such case isovector effects can be generated by a number of conventional mechanisms both at the nuclear and nucleon levels 
in addition to explicit isovector effects in the nuclear modification of bound nucleons~\cite{Kulagin:2014vsa}. 

The study of isospin symmetry violations in free nucleons and of isovector contributions to nuclear corrections is particularly relevant for 
long-baseline neutrino oscillation experiments using non-isoscalar nuclear targets like Ar~\cite{Abi:2020evt}. In this case the observation 
of CP violation in the leptonic sector relies on the detection of tiny differences between neutrino and antineutrino CC interactions, which 
are directly sensitive to isovector effects due to the relatively large non-isoscalarity of the target nucleus.

\section{Cross-sections for Free Nucleons} 
\label{sec:xsec} 

\begin{figure}[tb]
\begin{center}
\includegraphics[width=0.50\textwidth]{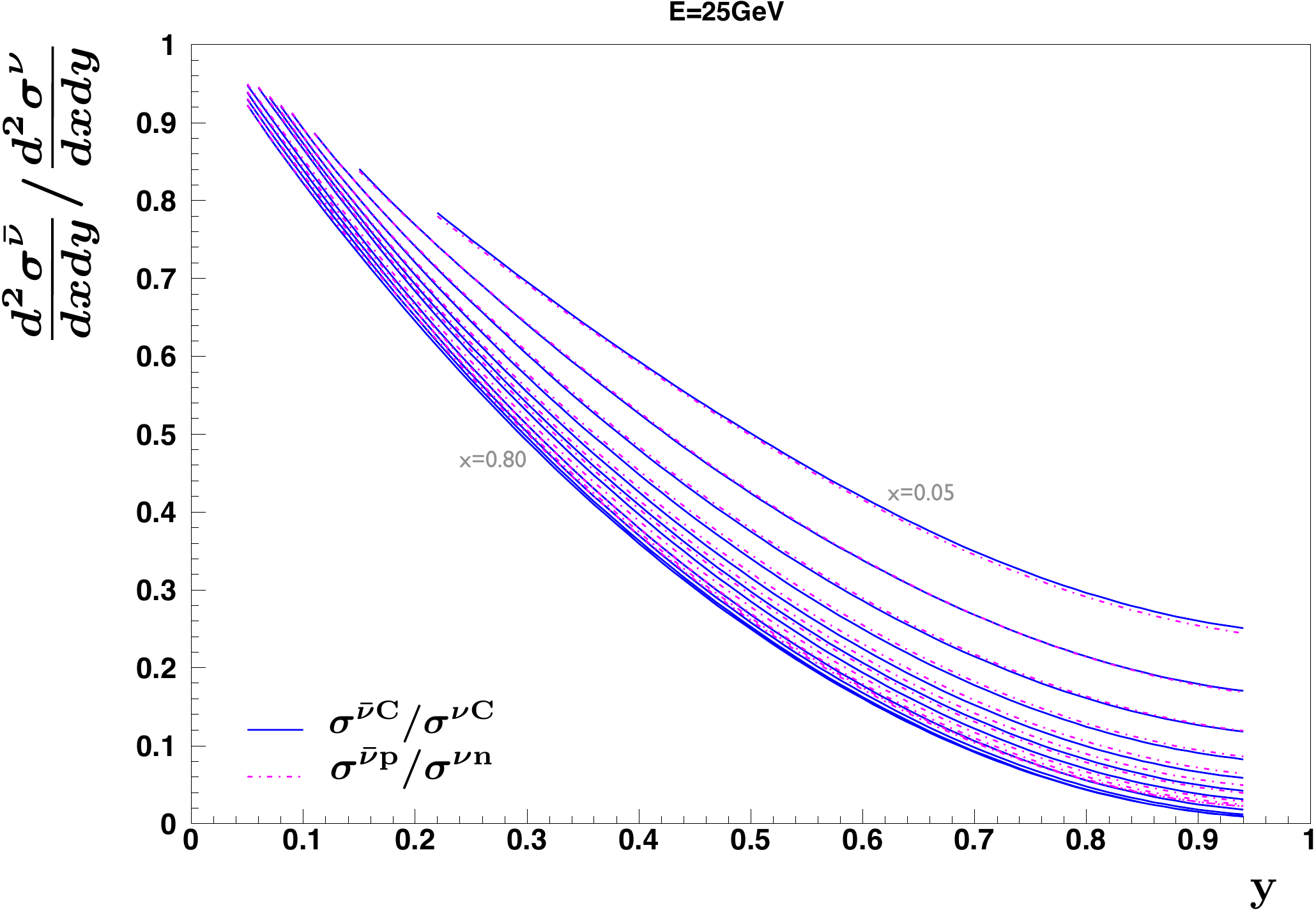}\hspace*{0.10cm}\includegraphics[width=0.50\textwidth]{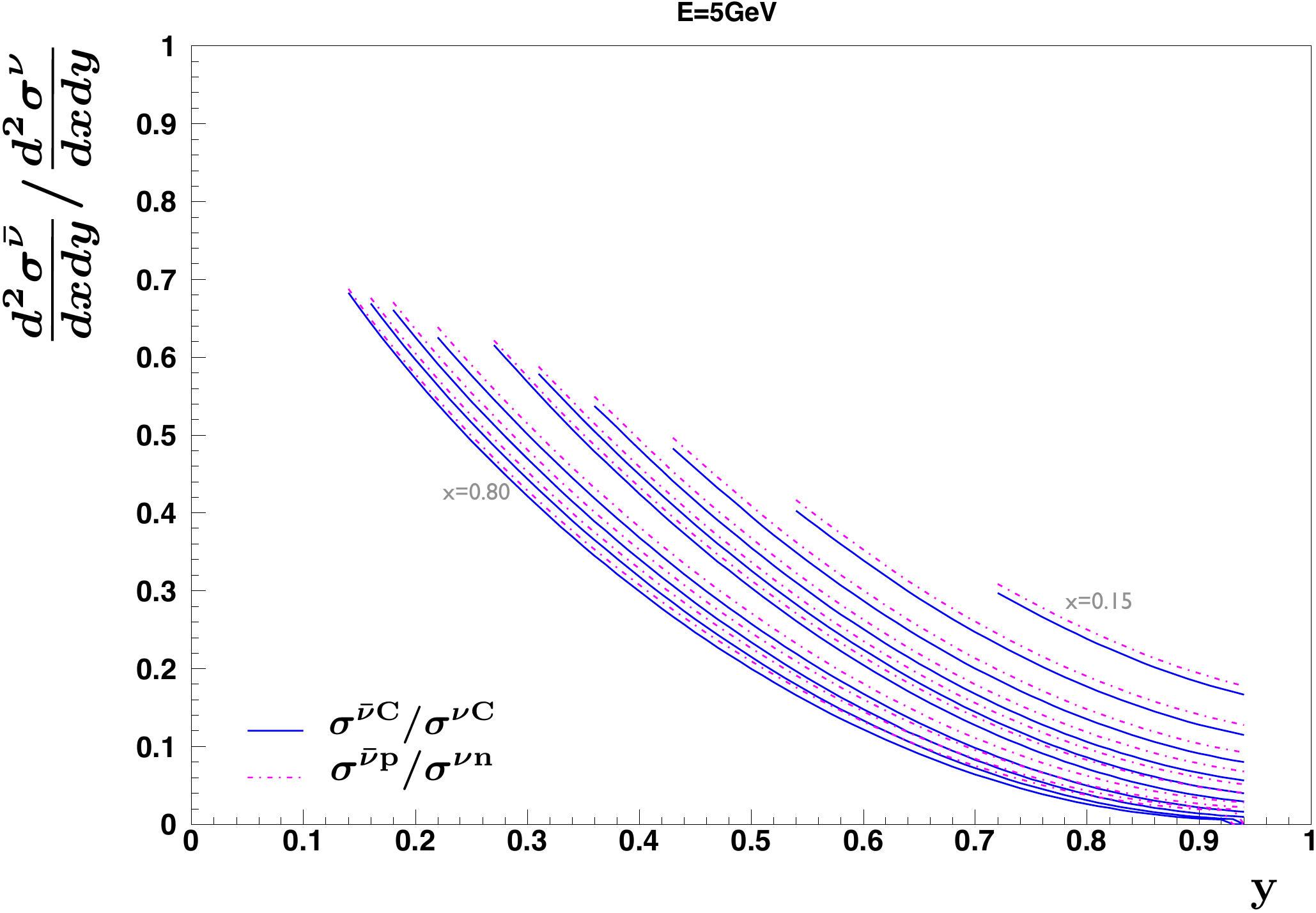}
\end{center}
\caption{Ratio between antineutrino and neutrino differential cross-sections $d^2 \sigma / dxdy$ as a function of the inelasticity $y$ for $Q^2 \geq 1$ GeV$^2/c^2$. 
A comparison between ${\cal R}_\sigma^{C}$ for an isoscalar C target (solid lines) and ${\cal R}_\sigma^{\rm p/n}$ for free nucleons (dash-dotted lines) is shown. 
The various curves correspond to values of $x=0.05,0.1,0.15,0.2,0.25,0.3,0.35,0.4,0.5,0.6,0.7,0.8$. 
See text for details. 
}
\label{fig:isospin-xsec}
\end{figure}

For studies related to the reconstruction of the neutrino energy and comparisons with data from nuclear targets it would be desirable to have 
physical events from a free neutron target, in addition to the structure function measurement discussed in Sec.~\ref{sec:neutron}. 
However, the parity violating axial-vector component of the weak current implies different kinematic factors for the $\nu$ and $\bar \nu$ 
cross-sections, resulting in large differences in the corresponding Bjorken $y$ distributions. We can quantify the impact of 
such differences by considering the following ratios between $\bar \nu$ and $\nu$ differential cross-sections: 
\begin{eqnarray}  
{\cal R}_\sigma^{\rm p/n}  (E_\nu,x,y) & = & \frac{d^2 \sigma^{\bar \nu p}}{dxdy} (E_\nu,x,y) \; / \; \frac{d^2 \sigma^{\nu n}}{dxdy} (E_\nu,x,y) \\ 
{\cal R}_\sigma^{C}  (E_\nu ,x,y) & = & \frac{d^2 \sigma^{\bar \nu C}}{dxdy} (E_\nu,x,y) \; / \;  \frac{d^2 \sigma^{\nu C}}{dxdy} (E_\nu,x,y)
\end{eqnarray} 
where $E_\nu$ is the neutrino energy. 
The ratio for free nucleons ${\cal R}_\sigma^{\rm p/n}$ is not directly measurable and represents the correction factor 
needed to use $\bar \nu$-H CC events as an approximation for $\nu n$ CC events. The ratio ${\cal R}_\sigma^{C}$ is the corresponding 
quantity for the isoscalar C target. Figure~\ref{fig:isospin-xsec} shows a comparison between the two ratios as a function of the 
inelasticity $y$ for different values of $x$ and the (anti)neutrino energy $E_\nu$. 
Similarly to what observed for the structure functions $F_2$ and $xF_3$ in Sec.~\ref{sec:neutron}, the results for the C target 
are relatively close to the ones for the free nucleons with the differences decreasing at higher energies. 
The C target can thus provide a control sample to validate the calculated ${\cal R}_\sigma^{\rm p/n}$, in addition to the 
structure function ratios ${\cal R}_2^{\rm p/n}$ and ${\cal R}_3^{\rm p/n}$. The ratio ${\cal R}_\sigma^{\rm p/n}$ could be used 
to re-weight the detected $\bar \nu$-H CC events to reproduce $\nu n$ CC events on free neutron.

\section{Smearing from Nuclear Targets} 
\label{sec:nuclear} 

The availability of both H and nuclear targets within the same detector allows the combined use of both $\nu$-H and $\bar \nu$-H CC interactions 
to calibrate the neutrino energy scale in CC interactions from the nuclear targets. As discussed in Sec.~\ref{sec:intro}, the problem arises because 
in conventional (anti)neutrino wide band beams the energy of the incoming neutrino is unknown on an event-by-event basis.
The need to infer the neutrino energy from the detected final state particles constitutes an intrinsic limitation of neutrino scattering experiments 
using nuclear targets, as the nuclear smearing introduces substantial systematic uncertainties in the process. 
The number of detected events originated from CC interactions with the generic nucleus $A$ can be written as: 
\begin{eqnarray}  
N^{\rm A} (E_{\rm rec}) & = & \int dE_\nu \Phi (E_\nu) \sigma^{\rm A}  (E_\nu) R_{\rm phys}^{\rm A} (E_\nu, E_{\rm vis}) R_{\rm det}^{\rm A} (E_{\rm vis}, E_{\rm rec}) 
\label{eq:Nevt} 
\end{eqnarray} 
where $\Phi$ is the input neutrino flux, $\sigma^{\rm A}$ the cross-sections for the process considered on the given nucleus, 
$R_{\rm phys}^{\rm A}$ the physics response function introduced by the nuclear smearing resulting in the visible final state particles, 
and $R_{\rm det}^{\rm A}$ is the detector response function (acceptance) for such particles. 
The smearing $R_{\rm phys}^{\rm A}$ is an irreducible effect of the target nucleus $A$ and is present even for an ideal detector. 
The variables $E_{\rm vis}$ and $E_{\rm rec}$ represent the total energy of the visible final state 
particles emerging from the nucleus and the final reconstructed energy in the detector, 
respectively. Similar equations can be written for any other kinematic variable, by simply replacing 
the energy with the corresponding variable. We note that the terms on the right side of \eq{eq:Nevt} are folded together 
into the observed event distributions and cannot be decoupled by using a single nuclear target. 

In order to address the main systematic uncertainties affecting neutrino scattering 
experiments~\footnote{The systematic uncertainties originated from each of the terms in \eq{eq:Nevt} are also relevant for 
long-baseline neutrino oscillation experiments, in which a distorted flux $\Phi (E_\nu) P_{\rm osc}(E_\nu)$ is expected and 
the oscillation probability $P_{\rm osc}$ has to be inferred from the number $N^{\rm A}$ of observed $\nu$ and $\bar \nu$ events off a nuclear target.} 
we would need to constrain each of the terms appearing in the integrand of \eq{eq:Nevt} with direct measurements using 
appropriate data control samples~\cite{Petti:2019asx}.  The flux $\Phi$ is the only term to be easily factored out. 
The relative $\nu_\mu$ and $\bar \nu_\mu$ flux as a function of energy can be 
determined in-situ with an accuracy better than 1\% using exclusive $\nu_\mu$ single pion and $\bar \nu_\mu$ quasi-elastic processes on H at 
small energy transfer~\cite{Duyang:2019prb}. The detector acceptance $R_{\rm det}^{\rm A}$ is controlled by the reconstruction 
efficiency and the energy scales of individual final state particles, which can be calibrated with an accuracy of 0.2\% using reconstructed 
$K_0 \to \pi^+\pi^-$ and $\Lambda \to p \pi^-$ decays in the detector considered~\cite{Wu:2007ab,Duyang:2019prb}. 
However, significant uncertainties are associated with the nuclear cross-section $\sigma^{\rm A}$~\cite{NuSTEC:2017hzk}, 
and the nuclear smearing $R_{\rm phys}^{\rm A}$ is essentially unknown. 
To this end, the $\nu$-H and $\bar \nu$-H CC interactions obtained with the ``solid" hydrogen technique 
can provide a calibration control sample free from nuclear effects. 
We note that an accurate knowledge of the input spectrum $\Phi (E_\nu)$ is not sufficient to fully constrain $R_{\rm phys}^{\rm A}$ 
from \eq{eq:Nevt} and to exclude potential degeneracies in the associated smearing matrix. 

\begin{figure}[tb]
\begin{center}
\includegraphics[width=0.65\textwidth]{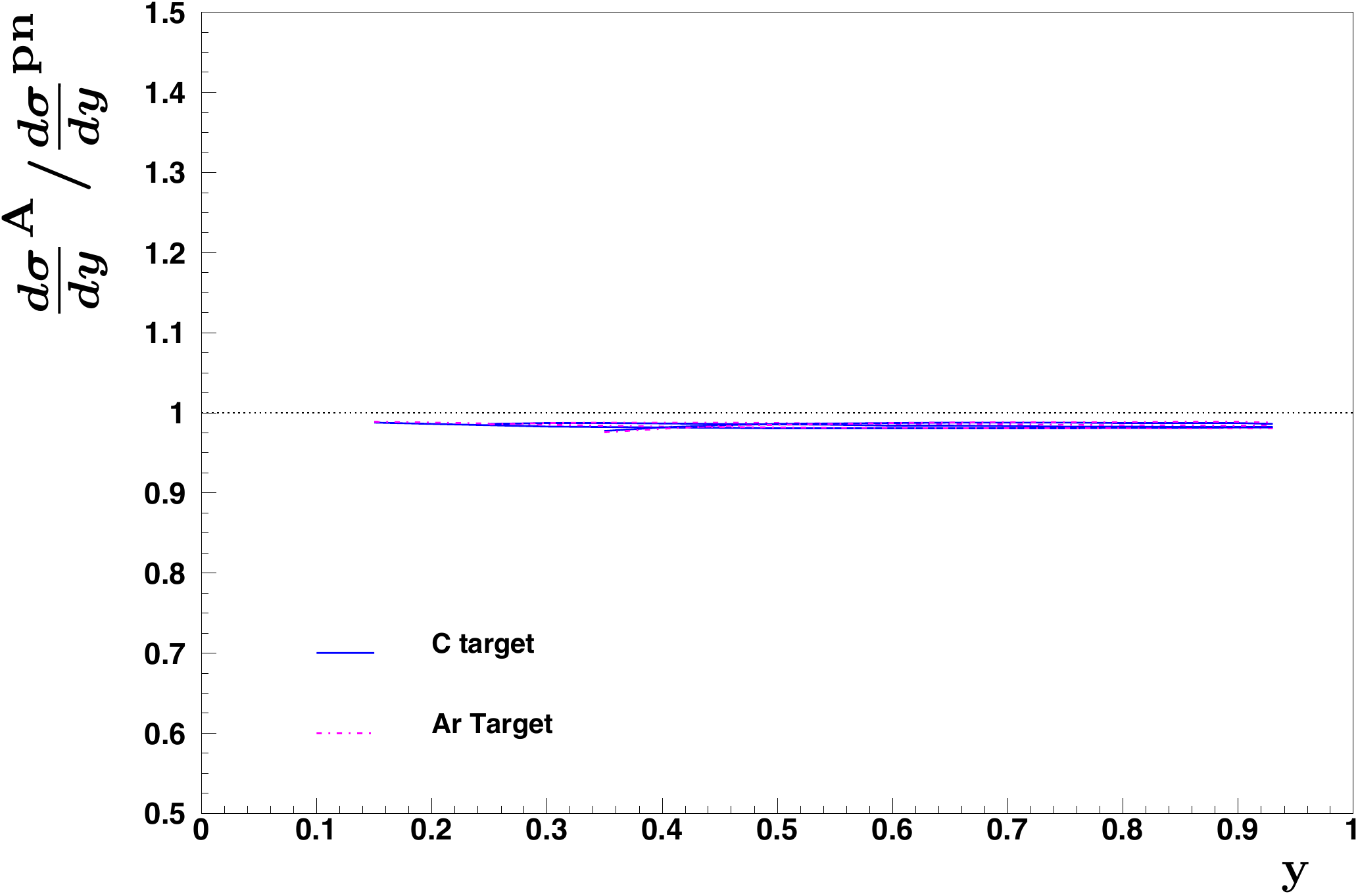}
\end{center}
\caption{Nuclear modification to the cross-section $d \sigma / dy$ integrated over $x$ in the range of the inelasticity $y$ covering $>90\%$ of the total 
cross-section for $E_\nu=5,10,25$ GeV. Results for both C (solid lines) and Ar (dash-dotted lines) targets are shown for comparison. See text for details.  
}
\label{fig:xsec-ratio}
\end{figure}

With a simple re-arrangement of the terms we can factorize nuclear effects from the interactions on an ideal target composed of $Z$ 
free protons and $(A-Z)$ free neutrons: 
\begin{eqnarray}  
N^{\rm A} (E_{\rm rec}) & = & \int dE_\nu \left[ \Phi (E_\nu)   \sigma^{\rm pn} (E_\nu) I(E_\nu, E_{\rm vis}) R_{\rm det}^{\rm pn} (E_{\rm vis}, E_{\rm rec}) \right] \times  \nonumber \\ 
     &  &  \;\; \;\; \left[  \frac{\sigma^{\rm A}}{\sigma^{\rm pn}}   (E_\nu) R_{\rm phys}^{\rm A} (E_\nu, E_{\rm vis}) \frac{R_{\rm det}^{\rm A}}{R_{\rm det}^{\rm pn}} (E_{\rm vis}, E_{\rm rec}) \right]  \nonumber \\ 
     & \equiv & \int dE_\nu  N^{\rm pn} (E_\nu, E_{\rm rec}) \times  R(E_\nu, E_{\rm rec}) 
\label{eq:fact} 
\end{eqnarray} 
where the index ${\rm pn}$ denotes quantities referring to the ideal free nucleon target, for which $R_{\rm phys}^{\rm pn}  = I$. The first term in the integrand is the number of events originated from interactions with the free nucleon target, $N^{\rm pn}$. The second term $R$ incorporates the nuclear modification to the cross-section, 
$\sigma^{\rm A} / \sigma^{\rm pn}$, the effect of the nuclear smearing on the neutrino energy, $R_{\rm phys}^{\rm A}$, and the ratio 
$R_{\rm det}^{\rm A} / R_{\rm det}^{\rm pn}$ between the detector acceptance for the nuclear target $A$ and the one for the free nucleon target ${\rm pn}$. 

The STT is designed to integrate a variety of thin targets within the tracking volume with the same acceptance 
for final state particles (Sec.~\ref{sec:proton}). In this case we expect $R_{\rm det}^{\rm A} / R_{\rm det}^{\rm pn} \simeq 1$ for inclusive CC interactions since the 
detector acceptance is dominated by the outgoing lepton and is less sensitive to the specific hadronic final states. 

Nuclear modifications to the (anti)neutrino cross-sections largely cancel out once integrated over $x$ and $Q^2$, due to 
the conservation of baryon number and DIS sum rules~\cite{Kulagin:2007ju}. 
This cancellation is illustrated in Fig.~\ref{fig:xsec-ratio} for the differential cross-section 
$d \sigma / dy$ as a function of the inelasticity $y$ and implies that the ratio $\sigma^{\rm A} / \sigma^{\rm pn} (E_\nu) \simeq 1$ 
away from the kinematic boundaries. We can therefore conclude that for the detector we are considering the second term in \eq{eq:fact} can 
approximate the nuclear smearing on $E_\nu$, i.e. $R \simeq R_{\rm phys}^{\rm A}$.

Assuming a discrete binning for both the $N^{\rm A}$ and $N^{\rm pn}$ distributions we can write: 
\begin{eqnarray}  
N^{\rm A}_j (E^\prime) & = & \sum_{i=1}^{K} N_i^{\rm pn} (E ) R_{ij} (E, E^\prime ) 
\label{eq:NevtA} 
\end{eqnarray} 
where $R_{ij} (E, E^\prime )$ is the smearing matrix describing the migration from the $E$ energy to the $E^\prime$ energy and 
the total number of events is the same in both distributions $\sum_{j=1}^{K} N^{\rm A}_j = \sum_{i=1}^{K} N_i^{\rm pn}$.

\section{\boldmath Calibration of Neutrino Energy with $\nu(\bar \nu)$-H} 
\label{sec:calibration} 

We can obtain $N_i^{\rm pn}$ from $\nu$-H and $\bar \nu$-H CC interactions using the isospin symmetry $\bar \nu p \leftrightarrow \nu n$, 
as discussed in Sec.~\ref{sec:xsec}: 
\begin{eqnarray} 
N_i^{\rm pn} (E - \Delta E) & = & \frac{Z}{A} N_i^{\nu p} (E - \Delta E) + \frac{(A-Z)}{A} N_i^{\nu n} (E - \Delta E) 
\label{eq:Nevtpn} 
\end{eqnarray} 
where $\Delta E  = E - E_0$ is the effect of the nuclear smearing in the bin $N_i$, corresponding to the energy shift $E \to E_0$ for a selected reconstructed 
energy $E^\prime=E_0$. We subtract $\Delta E$ from each bin and calculate the corresponding shifted value of the inelasticity $y$. 
After this subtraction all $N_i$ bins will be characterized by the same reconstructed energy $E_0$. 
We then consider the shifted $y$ distribution for the linear combination: 
\begin{eqnarray}  
\sum_{i=1}^{K} N_i^{\rm pn} (E - \Delta E) R_{ij} (E, E_0) 
\label{eq:Npn_lin} 
\end{eqnarray} 
and compare it with the corresponding one for $N^{\rm A}_j (E_0)$ obtained by selecting the reconstructed energy $E_0$, using a number of bins $\geq K$. 
The elements $i=1,...,K$ of the $R_{ij} (E_0, E)$ smearing matrix can be obtained by fitting the $y$ distribution for $N^{\rm A}_j$ with the linear combination. 
To this end, we can restrict the analysis to the range $0.1 \leq y \leq 0.9$ in order to avoid the regions characterized by larger electroweak 
corrections~\cite{Arbuzov:2004zr} and closer to the kinematic boundaries.

\begin{acknowledgments}
We thank S. Kulagin and S. Alekhin for fruitful discussions and collaboration on some of the topics covered. 
This work was supported by Grant No. DE-SC0010073 from the Department of Energy, USA.

\end{acknowledgments}

\bibliography{main1}

\end{document}